\documentclass[onecolumn,showpacs,preprintnumbers,amsmath,amssymb,showkeys,12ft]{revtex4}

\usepackage{graphicx}
\usepackage{dcolumn}
\usepackage{bm}

\begin{document}
\draft
\date{\today}
\title{Quantum Mechanical Treatment of the Problem of Constraints in Nonextensive Formalism Revisited }
\author{G. B. Ba\u{g}c\i $^{a}$, Altu\u{g} Arda$^{b}$, Ramazan Sever$^{c}$
}\thanks{Corresponding Author\\}
\email{sever@newton.physics.metu.edu.tr}
\address{$^{a}$ Department of Physics, University of North Texas, P.O. Box 311427, Denton, TX 76203-1427,
USA\\
 $^{b}$ Physics Education, Hacettepe University, 06532,\\
Ankara, Turkey\\
$^{c}$ Department of Physics, Middle East Technical University, 06531,\\
Ankara, Turkey}

\pagenumbering{arabic}

\begin{abstract}
The purity of Werner state in nonextensive formalism associated with
two different constraints has been calculated in a previous paper by
G. B. Bagci et al. [G. B. Bagci et al. , Int. J. Mod. Phys. {\bf
20}, 2085 (2006)]. Two different results have been obtained
corresponding to ordinary probability and escort probability. The
former has been shown to result in negative values thereby leading
authors to deduce the advantage of escort probabilities over
ordinary probabilities. However, these results have been only
derived for a limited interval of q values which lie between 0 and
1. In this paper, we solve the same problem for all values of
nonextensive index q by using a perturbative approach and show that
the simultaneous use of both types of constraints is necessary in
order to obtain the solution for whole spectrum of nonextensive
index. In this sense, the existence of these different constraints
in nonextensive formalism must not be seen as a deficiency in the
formalism but rather must be welcomed as a means of providing
solutions for all values of parameter q.
\end{abstract}

  \narrowtext
\newpage \setcounter{page}{1}
\maketitle

\section{\protect\bigskip Introduction}

\noindent \qquad A nonextensive generalization of the standard
Boltzmann-Gibbs (BG) entropy has been proposed by C. Tsallis in
1988$^{\text{1-4}}$. This new definition of entropy is given by

\begin{equation}
S_{q}=k\frac{1-\sum_{i=1}^{W}p_{i}^{q}}{q-1},
\end{equation}
where k is a positive constant which becomes the usual Boltzmann
constant in the limit q$\rightarrow 1$, p$_{\text{i }}$ is the
probability of the system in the ith microstate, W is \ the total
number of the configurations of the system.The entropic index q is a
real number, which characterizes the degree of nonextensivity as can
be seen from the following pseudo-additivity rule:

\begin{equation}
S_{q}(A+B)/k=[S_{q}(A)/k]+[S_{q}(B)/k]+(1-q)[S_{q}(A)/k][S_{q}(B)/k],
\end{equation}
where A and B are two independent systems i.e., p$_{ij}$(A+B)$=$p$_{i}(A)$p$%
_{j}(B)$. As q$\rightarrow 1$, the nonextensive entropy definition
in Eq. (1) becomes

\begin{equation}
S_{q\rightarrow 1}=-k_{B}\sum_{i=1}^{W}p_{i}\ln p_{i},
\end{equation}
which is the usual BG entropy. This means that the definition of
nonextensive entropy contains BG statistics as a special case. The
cases q
\mbox{$<$}%
1, q
\mbox{$>$}%
1 and $q=1$ correspond to superextensivity, subextensivity and
extensivity, respectively.

The nonextensive formalism has been used successfully to investigate
earthquakes$^{\text{5}}$, models of fracture roughness$^{\text{6}}$,
entropy production$^{\text{7}}$, voltage-gated ion
channels$^{\text{8}}$ and climatological models$^{\text{9}}$.

The outline of the paper is as follows: In Section II, we use
quantum divergences in the nonextensive formalism in order to
calculate the purity of Werner state within perturbative approach
and analyze the role of different intervals of the values of
nonextensive parameter q. We show that each of them provides us with
a different answer and is complementary of one another. We summarize
the results in Section III.

\section{Bregman Divergence versus Csisz\`{a}r Divergence}

\bigskip The physical meaning of divergence which is also called relative entropy in the literature is
the free energy difference in ordinary (extensive) case.
\begin{equation}
\text{\ }K(\rho \parallel \sigma )=Tr[\rho (\ln \rho -\ln \sigma )].
\end{equation}

Quantum divergence has the same physical meaning in nonextensive
formalism as in the ordinary case explained above. Additionally,the
nonextensive divergence of Csisz\'{a}r type$^{\text{10}}$ has been
used by Abe and Rajagopal to show an expected violation of the
second law if the nonextensive index q is not in the range
(0,2]$^{\text{11}}$.

\begin{equation}
\text{\ }I_{q}(\rho \parallel \sigma )=\frac{1}{q-1}\text{\
}[Tr(\rho ^{q}\sigma ^{1-q})-1].
\end{equation}

Recently, an alternative definition of relative entropy of Bregman
type$^{\text{12}}$ has been provided by J. Naudts$^{\text{13}}$ and
T. D. Frank$^{\text{14}}$ which reads

\begin{equation}
\text{\ }D_{q}(\rho \parallel \sigma )=\frac{1}{q-1}\text{\
}[Tr(\rho ^{q})-Tr(\rho \sigma ^{q-1})]-[Tr(\rho \sigma
^{q-1})-Tr(\sigma ^{q})].
\end{equation}

The quantum divergence is also used to calculate purity of states in
quantum information theory. Unfortunately, The Kullback-Leibler
divergence given by Eq. (4) works only when the contribution of
$\rho$ is smaller than $\sigma$. If one chooses $\sigma$ to be a
pure state, then the contribution of $\sigma$ is smaller than the
contribution of $\rho$. Therefore, the Kullback-Leibler divergence
cannot be used for this particular case. In a previous study, S.
Abe$^{\text{15}}$ made use of Eq. (5) in order to calculate the
purity of Werner state in the nonextensive framework. For this
purpose, he used Werner state$^{\text{16}}$ which is given by the
density matrix

\begin{equation}
\text{\ }\rho _{W}=F\mid \Psi ^{-}\rangle \langle \Psi ^{-}\mid+\frac{1-F}{3%
}(\mid \Psi ^{+}\rangle \langle \Psi ^{+}\mid +\mid \Phi ^{+}\rangle
\langle
\Phi ^{+}\mid +\mid \Phi ^{-}\rangle \langle \Phi ^{-}\mid ),\text{ \ \ }%
\frac{1}{4}\leq F\leq 1
\end{equation}

where

\begin{equation}
\mid \Psi ^{\pm }\rangle \equiv \frac{1}{\sqrt{2}}(\mid +-\rangle
\pm \mid -+\rangle ),
\end{equation}

and

\begin{equation}
\mid \Phi ^{\pm }\rangle \equiv \frac{1}{\sqrt{2}}(\mid ++\rangle
\pm \mid --\rangle ).
\end{equation}

\bigskip

F is the fidelity of $\rho _{W\text{ }}$with respect to the pure
reference state $\sigma =\mid \Psi ^{-}\rangle \langle \Psi ^{-}\mid
.$ For the time being, we restrict ourselves to the interval q$\in
(0,1)$ since Eq. (5) should not be too sensitive to small
eigenvalues of the matrices as in Ref. [15]. If we substitute Werner
state in order to find the degree of purity with respect to $\sigma
$ in Eq.(5), we obtain

\begin{equation}
\text{\ }I_{q}(\rho _{W}\parallel \mid \Psi ^{-}\rangle \langle \Psi
^{-}\mid )=\frac{1}{1-q}(1-F^{q}).
\end{equation}

This is the result already obtained by Abe in Ref. [15]. If we
consider the alternative definition of quantum divergence proposed
by Naudts and T. D. Frank in Eq. (6) and redo the above calculation
by substituting $\sigma $, as defined earlier, we obtain

\bigskip
\begin{equation}
\text{\ }D_{q}(\rho \parallel \sigma )=\frac{1}{q-1}\text{\ }[F^{q}+3(\frac{%
1-F}{3})^{q}-F]-(F-1).
\end{equation}

\bigskip

Obviously, Eq. (11) is different from Eq. (10). It leads to negative
values for q$\in (0,1)$ and F smaller than 1. This result has been
obtained in Ref. [17] where it has been deduced that the relative
entropy of Bregman type is not suitable to handle these kind of
situations, since it is not positive definite for q $\in (0,1)$.
Now, let us look closer at these two different relative entropy
expressions i.e., Eqs. (5) and (6) by trying to solve them within a
perturbative approach. The reason for this is to ensure that all
eigenvalues of $\sigma $ are different than zero. From now on, we do
not have to restrict ourselves to any particular interval of q as
long as it is not equal to 1. In order to do this, let us rewrite
$\sigma $ as

\begin{equation}
\sigma =(1-\epsilon )\mid \Psi ^{-}\rangle \langle \Psi ^{-}\mid +\frac{%
\epsilon }{3}(1-\mid \Psi ^{-}\rangle \langle \Psi ^{-}\mid ).
\end{equation}

This definition of  $\sigma $ corresponds to our earlier definition
when we set $\epsilon $ equal to zero. If we recalculate Eqs. (5)
and (6) now, we obtain

\begin{equation}
\text{\ }I_{q}(\rho _{W}\parallel \sigma )=\frac{1}{q-1}[(1-\epsilon
)^{q}F^{q}+\epsilon ^{1-q}(1-F)^{q}-1)],
\end{equation}

whereas for Frank-Naudts version, we have

\begin{eqnarray}
\text{\ }D_{q}(\rho _{W}\parallel \sigma )=\frac{1}{q-1}\text{\ [}%
F^{q}+3^{1-q}(1-F)^{q}-\\
\nonumber (1-\epsilon )^{q-1}F-(\frac{\epsilon }{3}%
)^{q-1}(1-F)]-(1-\epsilon )^{q-1}F-(\frac{\epsilon }{3})^{q-1}(1-F)+(1-%
\epsilon )^{q}+3^{1-q}\epsilon ^{q}.
\end{eqnarray}

Now let us consider the limit $\epsilon \rightarrow 0$ for these two
distinct expressions of divergence. Then, for q$\in (0,1)$, we
obtain

\begin{equation}
\text{\ }I_{q}(\rho _{W}\parallel \sigma )=\frac{1}{1-q}(1-F^{q})
\end{equation}
and
\begin{equation}
\text{\ }D_{q}(\rho _{W}\parallel \sigma )=+\infty .
\end{equation}
On the other hand, for q values greater than 1, we have
\begin{equation}
\text{\ }I_{q}(\rho _{W}\parallel \sigma )=+\infty
\end{equation}
and
\begin{equation}
\text{\ }D_{q}(\rho _{W}\parallel \sigma )=\frac{1}{q-1}\text{\ [}F^{q}+3(%
\frac{1-F}{3})^{q}-F]-(F-1).
\end{equation}

It is important to note that both type of quantum divergences share
the same feature of Kullback-Leibler measure in resulting
divergence. The Bregman type diverges for q values which lie between
0 and 1 whereas the Csisz\`{a}r type diverges for q values greater
than 1. But, we have non-diverging results due to the existence of
entropic index q which is an advantage of nonextensive formalism.
Note that the physical meaning of Bregman divergence is the same as
Eq. (4) i.e. difference of free energies if one employs ordinary
constraint$^{\text{18}}$ whereas the same meaning can be preserved
for Csisz\`{a}r type divergence by employing escort
probability$^{\text{19}}$ (see Ref. [20] for details).

\section{RESULTS AND DISCUSSIONS}

We have studied two definitions of divergence in current use in the
nonextensive formalism in order to calculate the degree of
purification of Werner state by adopting a perturbative method. This
in turn enabled us to generalize the results of our previous
work$^{\text{17}}$ for all values of the nonextensive index q. In
other words, the existence of two types of divergence associated
with two types of constraints makes the calculation possible for all
values of entropic index, rendering the problem of purity of state
fully solvable. In this sense, one can interpret the existence of
multiple constraints as an advantage of nonextensive formalism, not
a defect which is to be eliminated.

\section{ACKNOWLEDGEMENTS}
This research was partially supported by the Scientific and
Technological Research Council of Turkey. We also thank Jan Naudts
for pointing us the perturbative calculation included in this paper
and many insightful comments.

\end{document}